\documentclass[twocolumn]{article}
\usepackage[a4paper, total={7in, 10in}]{geometry}
\pdfoutput=1
\usepackage{graphicx}
\usepackage{bm}
\usepackage{amssymb,amsmath,latexsym}
\usepackage{color}
\usepackage[nottoc]{tocbibind}
\definecolor{darkblue}{RGB}{0,0,196}
\definecolor{darkred}{RGB}{190,45,30}
\definecolor{violet}{RGB}{155,38,182}
\definecolor{magenta}{RGB}{255,0,255}
\usepackage[colorlinks=true,linktocpage=true,linkcolor=darkblue,citecolor=darkblue,urlcolor=darkblue]{hyperref}
\usepackage{array}
\usepackage{subfigure}

\begin{document}

\title{Electromagnetic Debye mass within Gribov-Zwanziger action}
  
\author{Aritra Bandyopadhyay \\ email : \href{a.bandyopadhyay@thphys.uni-heidelberg.de}{a.bandyopadhyay@thphys.uni-heidelberg.de}}

\date{%
    Institut für Theoretische Physik, Universität Heidelberg, Philosophenweg 16, 69120 Heidelberg, Germany\\
}



\maketitle

\begin{abstract}
In the present study we have investigated the electromagnetic Debye mass by computing  the static limit of the temporal component of the one-loop photon polarization tensor involving effective quarks in the loop. These effective quarks have been considered within the Gribov-Zwanziger action, thereby incorporating the necessary non-perturbative effects. As an academic exercise, we have also shown one of the applications of this Gribov modified Debye mass by using it to evaluate the real and imaginary parts of the heavy quark potential in a QED system. This computation of the electromagnetic Debye mass and the corresponding estimations of the heavy quark potential can be seen as stepping stones in the direction of further practical QCD calculations in the non perturbative domain. 
\end{abstract}

\section{Introduction}

Due to the continuous influx of the experimental results from the ongoing ultra-relativistic heavy-ion collision experiments at RHIC and LHC, the existence of the deconfined quark matter at very high temperatures and/or densities is now a well known fact~\cite{Shuryak:1978ij,Collins:1974ky}. The characteristics of quark matter can be further explored by studying some important observables that serve as the signatures of quark matter. One such very important signature of quark matter is the set of n-point correlation functions. The fact that they are connected to a plethora of useful observables like dilepton production rate, transport coefficients and Debye screening mass, makes their investigation all the more essential in this work. We will however confine our interest mostly in exploring the electromagnetic (EM) Debye screening mass within a particular scenario as discussed in the following. 


A strongly coupled quark matter can only be explained using pure perturbative methods like Hard Thermal Loop (HTL) resummation~\cite{Braaten:1989mz,Braaten:1989kk} in the limit of very high temperature ($T>T_c$, $T_c \sim 160$ MeV being the pseudocritical temperature of the QCD phase diagram). In those relatively higher temperatures we have grasped enough in-depth understanding of the various collective modes originating from the quark matter using resummed perturbation theories (pt), e.g. HTLpt~\cite{Andersen:1999fw,Andersen:2002ey,Haque:2014rua}. These collective modes can be roughly classified into three types associated with three different thermal scales. Besides the energy (or hard) scale $T$ (temperature of the system), the screening of the electric fields are linked with the electric scale $gT$, where $g$ is the strong coupling. Moreover there is also the magnetic scale $g^2T$ connected to the behavior of magnetic fields in the plasma. The magnetic scale remains a significant challenge for theorists to handle systematically. While its inclusion removes infrared divergences, the physics associated with the magnetic scale is entirely non-perturbative. The magnetic sector is described by a dimensionally reduced three-dimensional Yang-Mills theory, and its non-perturbative nature is linked to the theory's confining properties. Earlier attempts to compute the non-abelian Debye screening mass faced issues, as a perturbative approach requires the electrostatic potential to acquire a gauge-symmetry-breaking vacuum expectation value, which leads to the breakdown of perturbation theory~\cite{PhysRevD.34.3904}. This shows that the study of the quark-gluon plasma is beyond a complete grasp of perturbation theory even at very high temperatures, making the inclusion of non-perturbative techniques essential. In recent years such efforts have been done using the Gribov-Zwanziger (GZ) action~\cite{Gribov:1977wm,Zwanziger:1989mf}, which takes into account the non-perturbative physics related to confinement associated with the magnetic scale $g^2T$ of the collective modes~\cite{Dokshitzer:2004ie,Sobreiro:2005ec,Vandersickel:2012tz}. GZ action introduces a new Gribov parameter $\gamma_G$ in the gluon propagator $D^{\mu\nu}(P)$, i.e.
\begin{align}
D^{\mu\nu}(P)=\left[\delta^{\mu\nu}-(1-\xi)\frac{P^\mu P^\nu}{P^2}\right]\frac{P^2}{P^4+\gamma_G^4}\, ,
\label{modified_gluon_prop}
\end{align}
which moves the poles of the gluon propagator to a nonphysical location ($P^2=\pm~ i\gamma_G^2$). Presence of this non-physical mass-like term within the gluon propagator leads to a suppression of long-wavelength gluons. The gluon propagator tends to zero in the infrared region, thereby prompting the screening of color charges at long distances and making the theory a confining one. The dimensionful Gribov parameter $\gamma_G$ acquires a well-defined interpretation if the topological structure of the SU(3) gauge group is aligned with the theory. By considering the periodicity of the theory’s $\theta$-vacuum, which arises from the compact nature of the SU(3) gauge group, $\gamma_G$ can be interpreted as the topological susceptibility of pure Yang-Mills theory~\cite{Kharzeev:2015xsa}, which also aligns with Zwanziger's original perspective that $\gamma_G$ is a statistical parameter~\cite{Zwanziger:1989mf}. In practice, $\gamma_G$ can be self-consistently determined using a one-loop gap equation, and at asymptotically high temperatures, a perturbative form of the parameter $\gamma_G$ up to one loop order and for a four-dimensional theory can be given as~\cite{Zwanziger:2006sc}
\begin{align}
\gamma_G = \frac{3N_c}{16\sqrt{2}\pi}g^2T,  \label{Gribov_para}
\end{align}
where $N_c$ represents the number of colors. This modified gluon propagator within GZ action has already been used in several avenues of exploration to study various properties of quark matter. These studies have produced various exciting results in several fronts, e.g. Yang-Mills thermodynamics~\cite{Fukushima:2013xsa}, quark collective modes~\cite{Su:2014rma}, dilepton production rate~\cite{Bandyopadhyay:2015wua}, transport properties~\cite{Jaiswal:2020qmj}, heavy quark dynamics~\cite{Madni:2022bea} etc. More recently, a refined GZ approach has surfaced which has also produced some interesting results~\cite{Dudal:2008sp,Capri:2016aqq,Dudal:2017kxb,Gotsman:2020mpg,Gotsman:2020ryd}. The in-medium quark propagation within GZ action generates a novel massless spacelike collective mode aside from the usual quark and plasmino modes, termed as the Gribov mode~\cite{Su:2014rma}. Appearance of this novel mode demands further analysis of the necessary modifications in the various signatures of the quark matter, some of which have already been done~\cite{Bandyopadhyay:2015wua}. In the present work we have studied the EM Debye mass using the effective quark propagator which includes this novel Gribov mode and report its consequences.

Static electric fields get screened at large distances because of the interaction between electrically charged particle and electromagnetic fields in quark matter. EM Debye mass ($m_D$) is basically the inverse screening length which is related with the corresponding vanishing of the static potential~\cite{Gross:1980br,Kajantie:1981hu,Kapusta:1989tk}. As an important spectral property, both EM and QCD Debye mass have also been studied within various scenarios in the presence of an external electromagnetic~\cite{Alexandre:2000jc,Bandyopadhyay:2016fyd,Bandyopadhyay:2017cle,Ayala:2018ina,Koothottil:2020riy} and gravitational~\cite{Popov:2017xut} fields. Moreover, the QCD Debye mass is significant in the context of the non-perturbative physics related to the magnetic scale associated with the GZ action. Both using resummed perturbation theory techniques and lattice QCD calculations when people extracted a complete and gauge-independent result for the non-abelian Debye screening mass at next-to-leading order from the static gluon propagator, the correction to the Debye mass is found to be logarithmically sensitive to the non-perturbative magnetic mass~\cite{Rebhan:1993az,Rebhan:1994mx,Braaten:1994qx,PhysRevD.52.7208,Kajantie:1997pd}. To the best of our knowledge, the present study incorporates GZ effective quark propagator in the computation of the Debye mass for the first time. As an application of the EM Debye mass computed in this work, we have also estimated the real and imaginary parts of the heavy quark potential in a QED system, where the in-medium effects are assumed to solely reside in the Debye mass. Heavy quarks (HQ) are also crucial in the context of characterizing quark matter as they interact less within the medium because of their relatively large masses. Discovery of the bound states like $J/\psi~(c\bar{c})$ and $\Upsilon~(b\bar{b})$ have prompted several interesting studies of the HQ potential under various circumstances~\cite{Karsch:1987pv,Matsui:1986dk,Alford:2013jva,Rougemont:2014efa,Singh:2017nfa,Hasan:2017fmf,Thakur:2013nia,Thakur:2020ifi,Agotiya:2016bqr,Jamal:2018mog,Burnier:2009yu,Nilima:2022tmz,Ghosh:2022sxi}, much of which have essentially been translated into distinctive refinements of the Debye mass. Hence the effect of a Gribov modified Debye mass on the HQ potential is certainly compelling to say the least, as recently showed in Ref~\cite{Debnath:2023dhs} where the authors have considered one-loop effective static gluon propagator. Finally we must emphasise here that the present estimations, being the first of its kind, is to an extent simplified and rudimentary in some aspects. But this computations should act as a stepping stone for further, more involved QCD calculations in near future, which would serve better practical purposes.

This paper is organized as follows. In section~\ref{sec2} we have computed the electromagnetic (EM) Debye mass with GZ modified effective quark propagator and discussed corresponding results. Next in section~\ref{sec3}, we have explored the effect of GZ modified quark propagation on the HQ potential, as one of the important applications of the EM Debye mass. Finally we summarize our results in section~\ref{sec4}.

\section{EM Debye Mass with GZ action}
\label{sec2}

\begin{figure}[h]
\begin{center}
\includegraphics[scale=1]{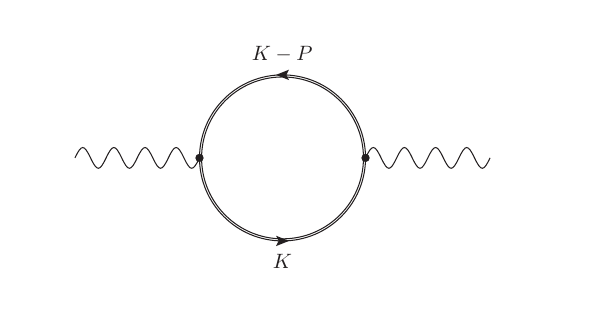}  
\end{center}
\caption{Photon self-energy diagram with GZ modified quarks in the loop (denoted by the double lines).}
\label{feyn_diag}
\end{figure}

The electromagnetic Debye screening mass is a direct consequence of thermal fluctuations, which drive the random motion of charged particles in a QED plasma. These fluctuations enable the plasma to effectively screen electric fields over a characteristic distance, the Debye screening length, which depends on the medium. As the temperature rises, particles move more rapidly, thereby enhancing the screening. Being related to the vanishing static potential, the electromagnetic Debye mass can be determined from the poles of the effective photon propagator at zero frequency (also known as the static limit). In the static limit, the electromagnetic Debye screening mass is defined as
\begin{align}
    m_D^2=\Pi_{00}(\omega=0,p\rightarrow 0),
\end{align}
where $\Pi_{00}$ is the temporal part of the photon self energy $\Pi_{\mu\nu}$. At one-loop order, the photon self energy diagram is shown in Fig.~\ref{feyn_diag}, which can be written as 
\begin{align}
\Pi_{\mu\nu}(P) = -i\sum_{f} q_f^2\int \frac{d^4K}{(2\pi)^4} 
\textsf{Tr}_{c}\left[\gamma_\mu S(K) 
\gamma_\nu S(Q)\right] ,
\label{photon_se}
\end{align}
where $Q=K-P$. The corresponding leading order result for the electromagnetic Debye mass yields - 
\begin{align}
    m_D^b = \frac{eT}{\sqrt{3}} + \mathcal{O}(e^2T),
    \label{Debyem_bare}
\end{align}
where superscript $b$ denotes the bare nature of the result.

For the present work, we will be using the GZ modified effective propagator $S^*(K/Q)$ in Eq.~\eqref{photon_se} (see Fig.~\ref{feyn_diag}). Using the helicity representation, the GZ modified effective fermion propagator can be written as~\cite{Su:2014rma,Bandyopadhyay:2015wua} 
\begin{align}
i S^*(K) &= \frac{1}{2} \frac{(\gamma_0 - \vec{\gamma}\cdot \hat{\bf{k}})}{D_+}
+ \frac{1}{2} \frac{(\gamma_0 + \vec{\gamma}\cdot \hat{\bf{k}})}{D_-} , \label{hprop}
\end{align}
where $D_\pm$ are explicitly given as
\begin{align}
D_+ &= k_0 - k - \frac{2g^2C_F}{(2\pi)^2}\sum_\pm\int dr~ r~ \tilde{n}_\pm(r,\gamma_G)  \nonumber \\
&\times \left[Q_0(\tilde{\omega}_1^\pm,k)+ Q_1(\tilde{\omega}_1^\pm,k)
+Q_0(\tilde{\omega}_2^\pm,k)+ Q_1(\tilde{\omega}_2^\pm,k)\right] ,\\
D_- &= k_0 + k - \frac{2g^2C_F}{(2\pi)^2}\sum_\pm\int dr~ r~ \tilde{n}_\pm(r,\gamma_G) \nonumber \\
&\times \left[Q_0(\tilde{\omega}_1^\pm,k)- Q_1(\tilde{\omega}_1^\pm,k)
+Q_0(\tilde{\omega}_2^\pm,k)- Q_1(\tilde{\omega}_2^\pm,k)\right]. 
\label{dpm}
\end{align}
Here $r$ is the loop momentum in the quark self energy and the modified frequencies are defined as $\tilde{\omega}_1^\pm \equiv E_\pm^0(k_0+r-E_\pm^0)/r$ and 
$\tilde{\omega}_2^\pm \equiv E_\pm^0(k_0-r+E_\pm^0)/r$ with $E_\pm^0 = \sqrt{r^2 \pm i\gamma_G^2}$. Modified distribution functions are given by $\tilde{n}_\pm(r,\gamma_G)\equiv n_B\!\left(\sqrt{r^2 \pm i\gamma_G^2}\right)+n_F(r)$, $n_B$ and $n_F$ being the Bose-Einstein and Fermi-Dirac distribution functions respectively. $Q_0$ and $Q_1$ are the Legendre functions of the second kind, defined as 
\begin{align}
Q_0(k_0,k) &\equiv \frac{1}{2k}\ln \frac{k_0+k}{k_0-k}, \\
Q_1(k_0,k) &\equiv \frac{1}{k}(1-k_0 Q_0(k_0,k)). \label{legd}
\end{align}

\begin{figure*}
    \centering
    \includegraphics[scale=0.38]{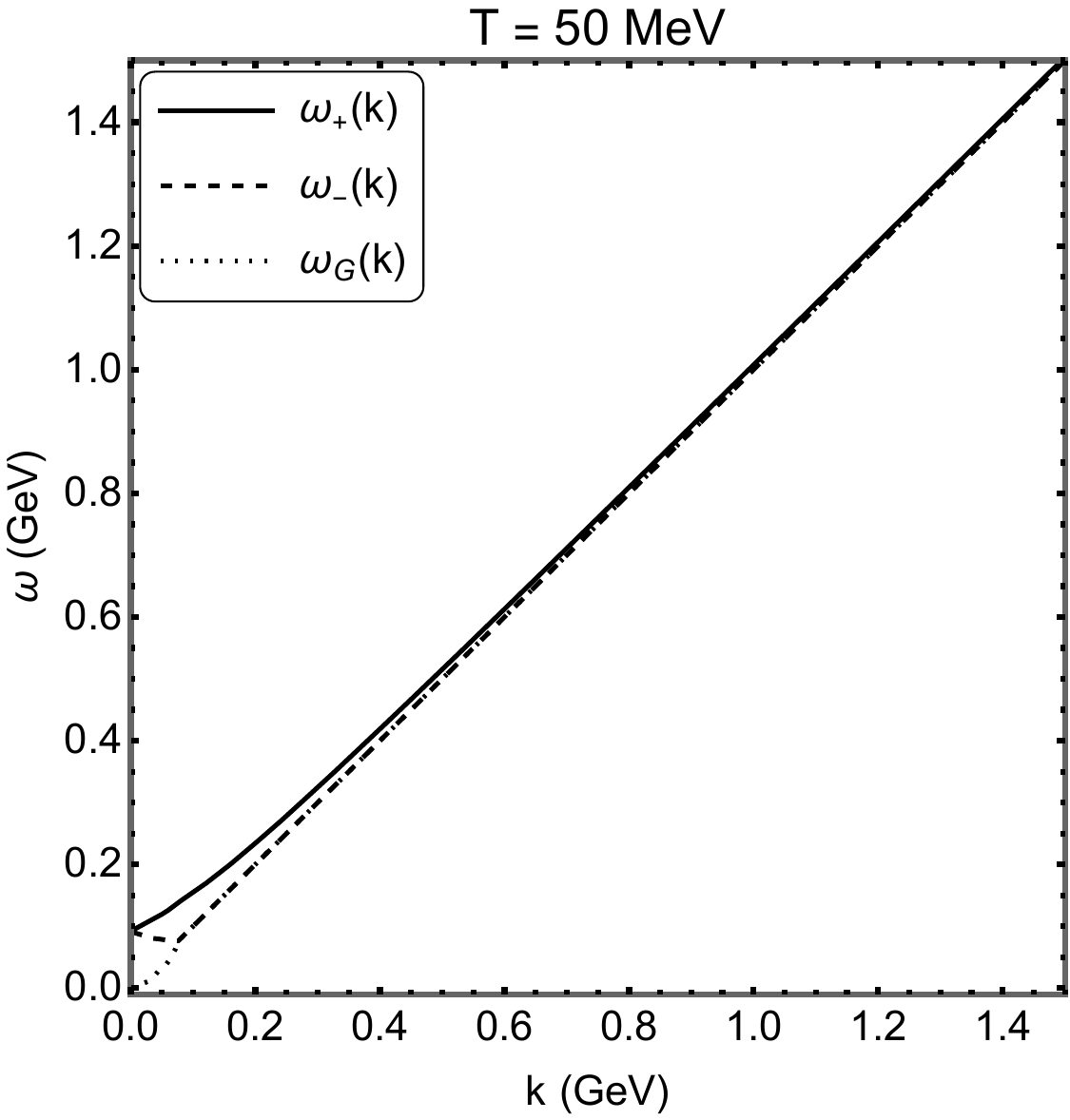}
    \hspace{0.6cm}
    \includegraphics[scale=0.38]{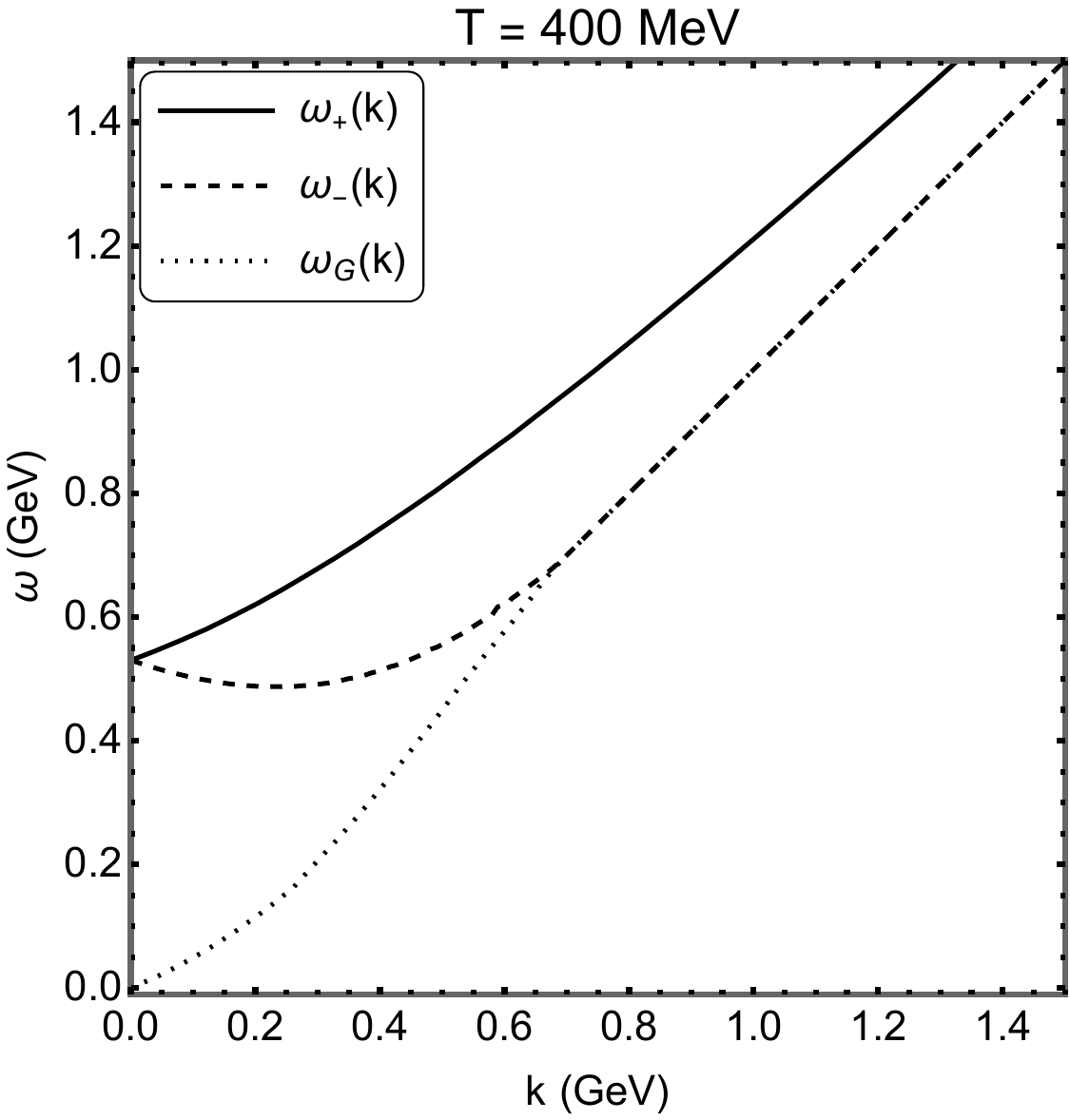}
    \caption{Three different poles of the GZ modified quark propagator are shown for two different temperatures, two extremes within the temperature range we have worked in the present study. These plots show the significant temperature dependence of the different poles.}
    \label{fig:disp_gribov}
\end{figure*}

As it would be of relevance later in our evaluation, we would also like to write down the corresponding spectral functions for $D_\pm$, which can be expressed as~\cite{Bandyopadhyay:2015wua}
\begin{align}
\rho_\pm^G(k_0,k) &= \frac{k_0^2-k^2}{2m_q^2(\gamma_G)}\left[\delta(k_0\mp\omega_+)+\delta(k_0\pm\omega_-)\right.\nonumber\\
&\left.+\delta(k_0\pm\omega_G)\right], 
\label{gspect}
\end{align}
where  $D_+$ has poles at $\omega_+$, $-\omega_-$, and $-\omega_G$ 
and $D_-$ has poles at $\omega_-$, $-\omega_+$, and $\omega_G$ with a prefactor,  $(\omega^2-p^2)/2m_q^2(\gamma_G)$ as the residue respectively. $m_q(\gamma_G)$ is the Gribov modified thermal quark mass, given as~\cite{Su:2014rma,Bandyopadhyay:2015wua}
\begin{align}
m_q^2(\gamma_G) = \frac{g^2C_F}{4\pi^2}\sum_\pm\int\limits_0^\infty dr \, \frac{r^2}{E_\pm^0} \, \tilde{n}_\pm(r,\gamma_G).
\label{tmass}
\end{align}
One can immediately notice that the spectral function that appears in Eq.~\eqref{gspect} has no Landau cut contribution, unlike usual HTL spectral functions. This is because $D_\pm$ posses no discontinuities at spacelike momenta in the complex plane.  Physically this signifies the absence of the high temperature quasigluons, which are usually incorporated through the perturbative methods. 

With these necessary information in our hand, we can now proceed to evaluate the temporal component of the photon self energy as 
\begin{align}
    \Pi_{00} = -i\sum_{f} q_f^2\int \frac{d^4K}{(2\pi)^4} 
\textsf{Tr}_{c}\left[\gamma_0 S^*(K)
\gamma_0 S^*(Q)\right].
\label{Pi00gribov_1}
\end{align}
After performing the color and Dirac traces, Eq.~\eqref{Pi00gribov_1} yields 
\begin{align}
    \Pi_{00} &= iN_c\sum_{f} q_f^2\int \frac{d^4K}{(2\pi)^4} 
\left[\frac{1+\hat{\bf k}\cdot\hat{\bf q}}{D_+(K)D_+(Q)}+\right.\nonumber\\
&\left.\frac{1+\hat{\bf k}\cdot\hat{\bf q}}{D_-(K)D_-(Q)}+\frac{1-\hat{\bf k}\cdot\hat{\bf q}}{D_+(K)D_-(Q)}+\frac{1-\hat{\bf k}\cdot\hat{\bf q}}{D_-(K)D_+(Q)}\right].
\end{align}

Implementing the static limit further simplifies the expression as, i.e.
\begin{align}
    m_D^2 &= \Pi_{00}(\omega=0,p\rightarrow 0) \nonumber \\
    &= 2iN_c\sum_{f} q_f^2\int \frac{d^4K}{(2\pi)^4} 
\left[\frac{1}{D_+(K)^2}+\frac{1}{D_-(K)^2}\right],\nonumber\\
&= -2N_c\sum_{f} q_f^2 T\sum_{k_0}\int\frac{d^3k}{(2\pi)^3} 
\left[\frac{1}{D_+(K)^2}+\frac{1}{D_-(K)^2}\right].
\end{align}

To perform the Matsubara frequency sum, we will use the spectral representation of the corresponding propagators, i.e. $D_\pm$. This process gives the following expressions for the frequency sums :
\begin{align}
    T\sum_{k_0} \frac{1}{D_\pm(K)^2} &= \int\limits_{-\infty}^{\infty} dk_0\frac{1-2n_F(k_0)}{2k_0}\rho^G_\pm(K)^2,
\end{align}
where $n_F$ is the Fermi-Dirac distribution function. Further, using this template for the frequency sum with the spectral function given in Eq.~\eqref{gspect}, we get the final expression for the EM Debye mass as : 
\begin{align}
     m_D^2 &= 2N_c\sum_{f} q_f^2 \int\frac{d^3k}{(2\pi)^3} \nonumber\\
     & \sum_{\omega=\omega_\pm,\omega_G}
\left[\left(\frac{\omega^2-k^2}{2m_q^2(\gamma_G)}\right)^2\frac{n_F(\omega)}{\omega}\right].
\label{Debyem_final}
\end{align}

One can notice that the modified EM Debye mass receives contributions from all the three individual collective modes of excitation originating from the quark propagation within a Gribov modified medium.  Eq.~\eqref{Debyem_final} is the primary analytic result of the present work which we will explore further in the following. With Eq.~\eqref{Debyem_final} in our hand, we move on to numerically evaluate the electromagnetic Debye mass, for which we need the values of different poles as functions of temperature. For each value of the temperature we solve for the zeroes of $D_\pm$ to get the corresponding $T$ dependent values of the poles, i.e. $\omega_+(k), \omega_-(k), \omega_G(k)$. As a demonstration, in Fig.~\ref{fig:disp_gribov} we present these poles for two extreme values of temperatures we have worked with, i.e. $T=50$ MeV and $T=400$ MeV. One can notice the drastic variation of the poles for different temperatures. For lower temperatures, both the Gribov mode $\omega_G$ (dotted curve) and the plasmino mode $\omega_-$ (dashed curve) rapidly approach the free mass-less propagation (not separately shown in Fig.~\ref{fig:disp_gribov}) and eventually merges with it simultaneously. So the medium effects for the lower regime of temperatures are confined within the very low momentum limit. Whereas, for relatively larger temperatures like $T=400$ MeV, the Gribov and plasmino modes merge with the free mass-less propagation at relatively much higher values of the momentum, i.e. $k\sim 0.7$ GeV. This implies that at higher values of temperatures the effect of incorporating the GZ action in any observable will be much more significant than that of the lower values of temperatures. This observation of enhanced effects of Gribov mode with increasing temperatures is really interesting since the excluded perturbative Landau cut contribution also contributes at relatively higher temperatures due to the quasigluons captured in the process. If one combines these two facts, it appears to be physically plausible that the Gribov mode contribution is compensating for the excluded Landau cut contribution.

\begin{figure}
    \centering
    \includegraphics[scale=0.4]{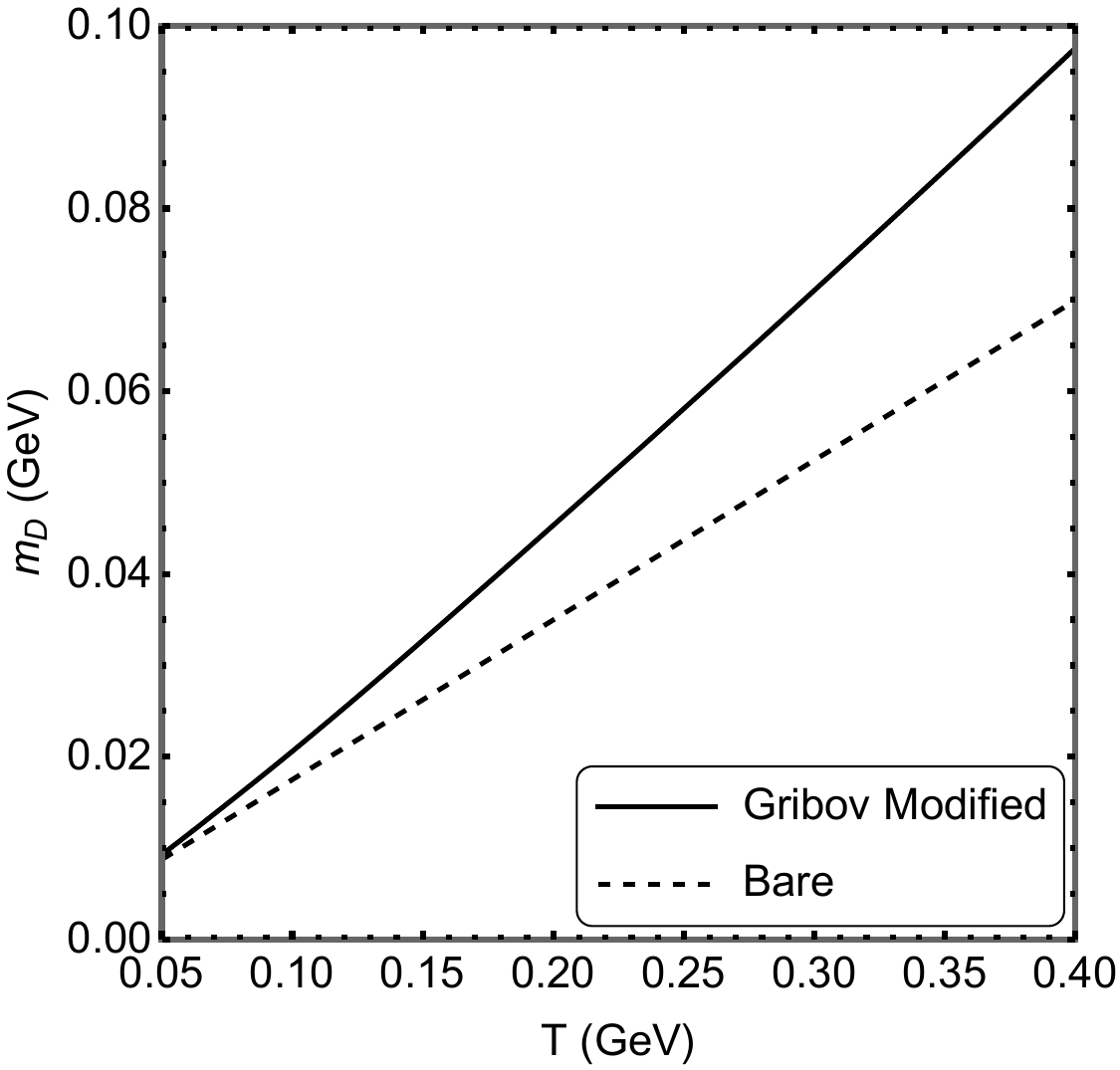}
    \caption{Comparison between the bare and the GZ modified Debye mass while varying with temperature}
    \label{fig:mD_gribov}
\end{figure}

\begin{figure*}
    \centering
    \includegraphics[scale=0.38]{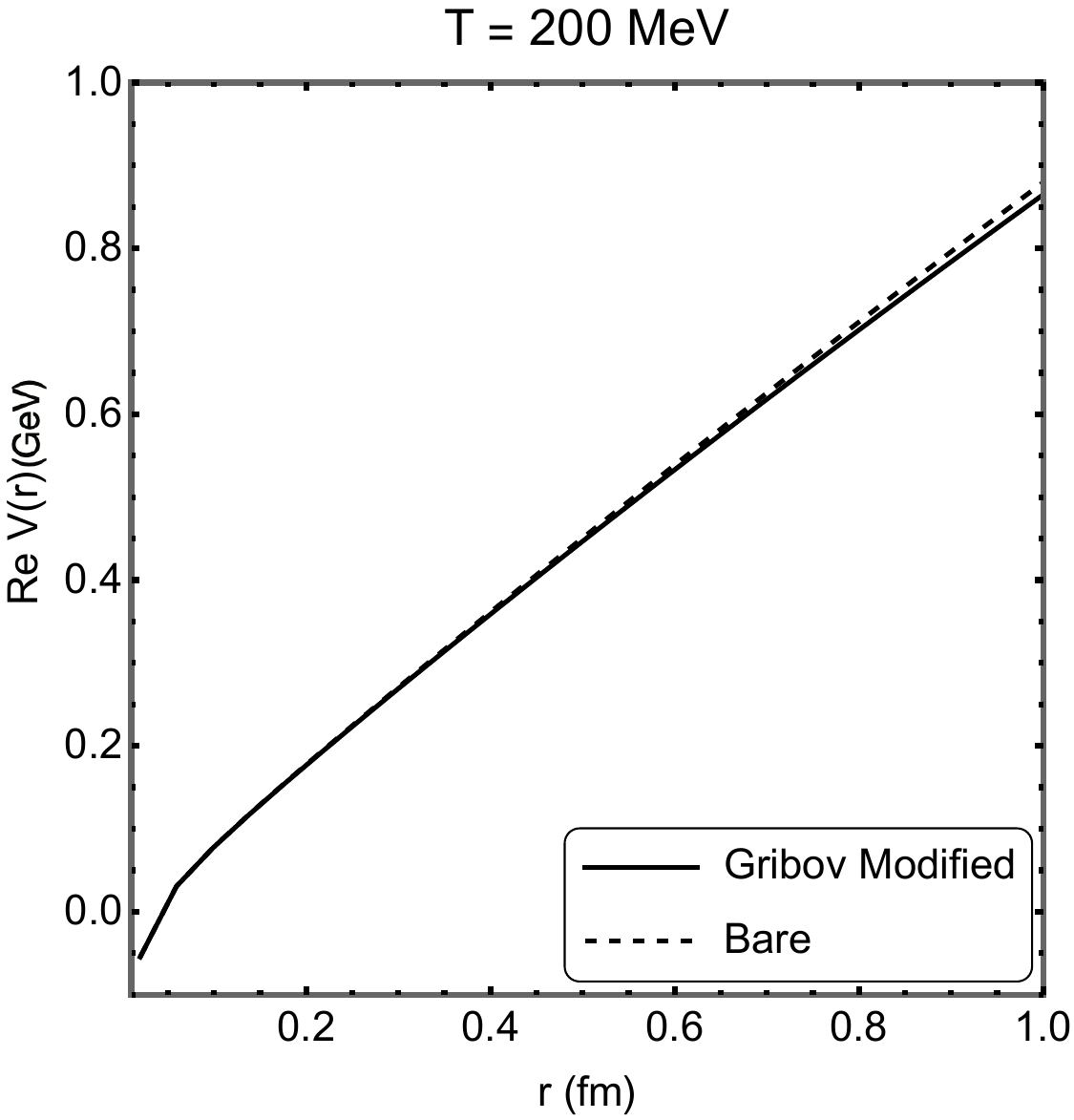}
    \hspace{0.6cm}
    \includegraphics[scale=0.38]{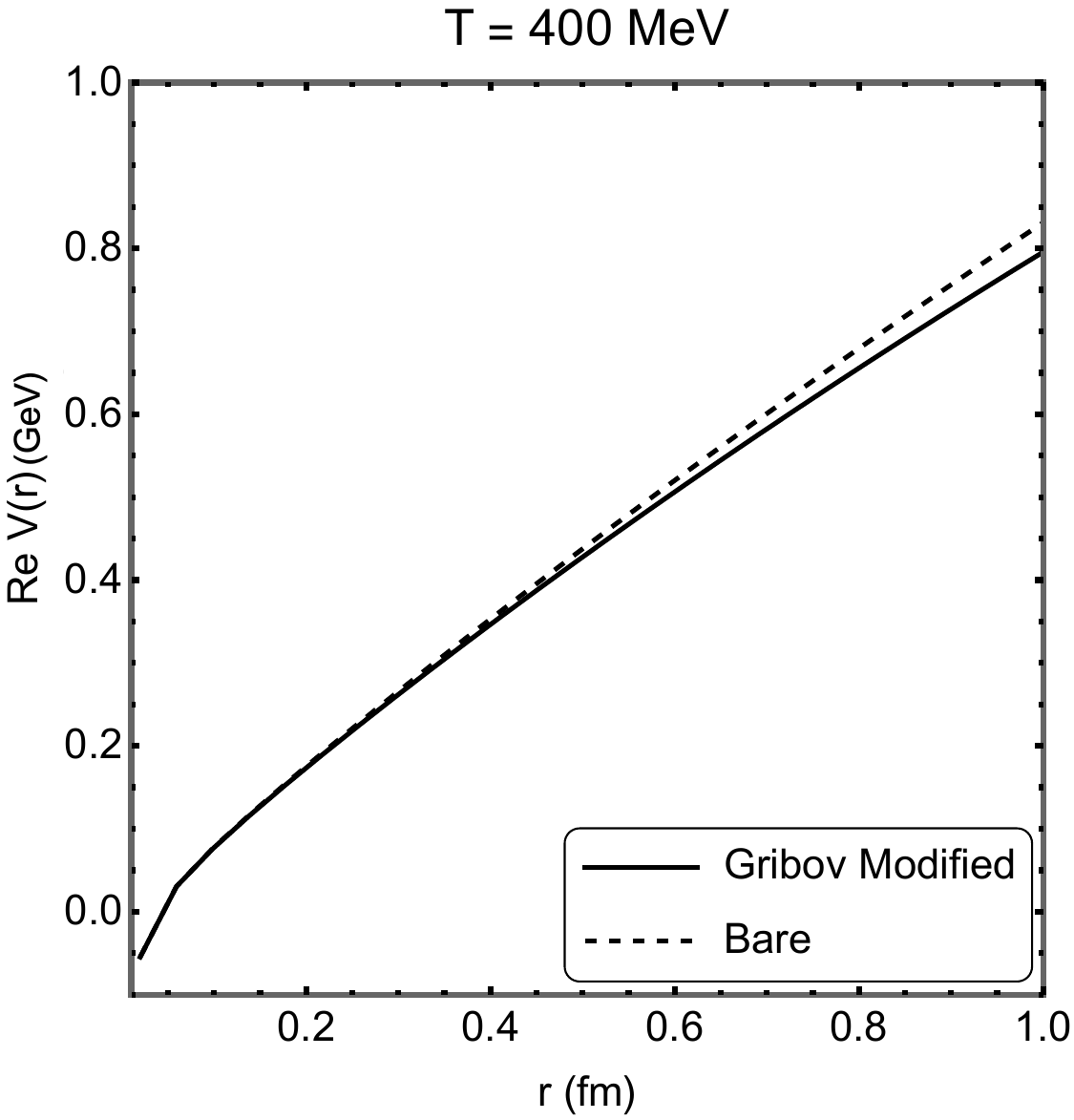}
    \caption{Comparison between the bare and the GZ modified real part of the HQ potential while varying with distance.}
    \label{fig:ReV_gribov}
\end{figure*}

Once we get the values of the $\omega_\pm(k)$ and $\omega_G(k)$ for different values of temperature, it is relatively straightforward to numerically evaluate the electromagnetic Debye mass as a function of the temperature using Eq.~\eqref{Debyem_final}. In Fig.~\ref{fig:mD_gribov} we present the main result of this study, where we compare the GZ modified electromagnetic Debye mass ($m_D$) with the bare result ($m_D^b$), given in Eq.~\eqref{Debyem_bare}. We have limited our comparison to $T$ as low as $0.05$ GeV, keeping in mind that the perturbative value of $\gamma_G$ used for our purpose has been obtained within the assumption of asymptotically high temperatures. It can be inferred from Fig.~\ref{fig:mD_gribov}, that although at lower temperatures there is not much difference between the bare and the Gribov modified results, at higher temperatures the GZ modified electromagnetic Debye mass clearly dominates the corresponding bare result. This result is also in accordance with the behavior noticed in Fig.~\ref{fig:disp_gribov}, which showed the enhanced significance of the Gribov and plasmino modes with increasing temperature. For high temperatures, the considerable enhancement of the GZ modified result over the bare result is the most important takeaway in the context of our present study.

\section{Application to Heavy Quark potential}
\label{sec3}

\begin{figure*}
    \centering
    \includegraphics[scale=0.38]{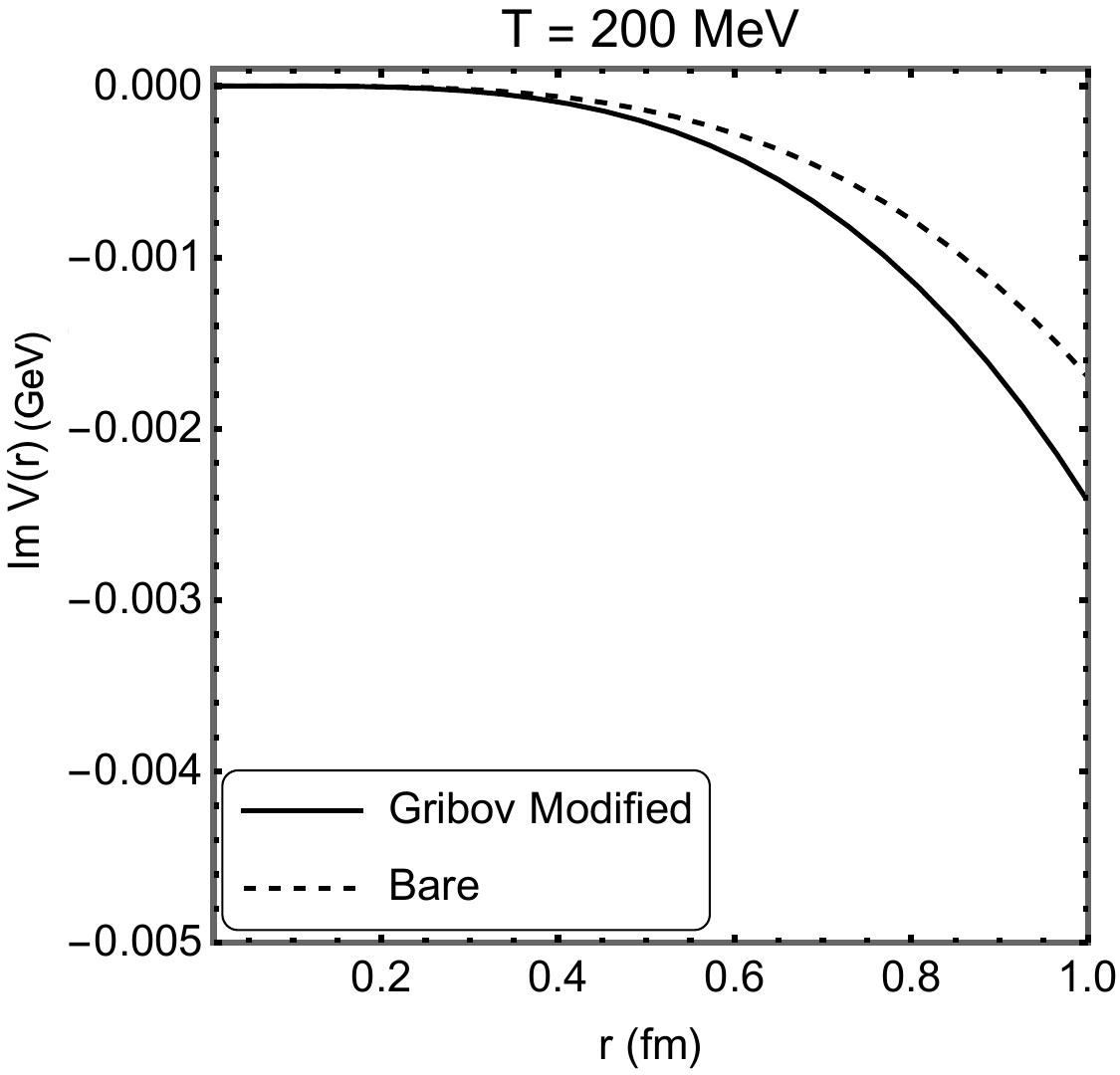}
    \hspace{0.6cm}
    \includegraphics[scale=0.38]{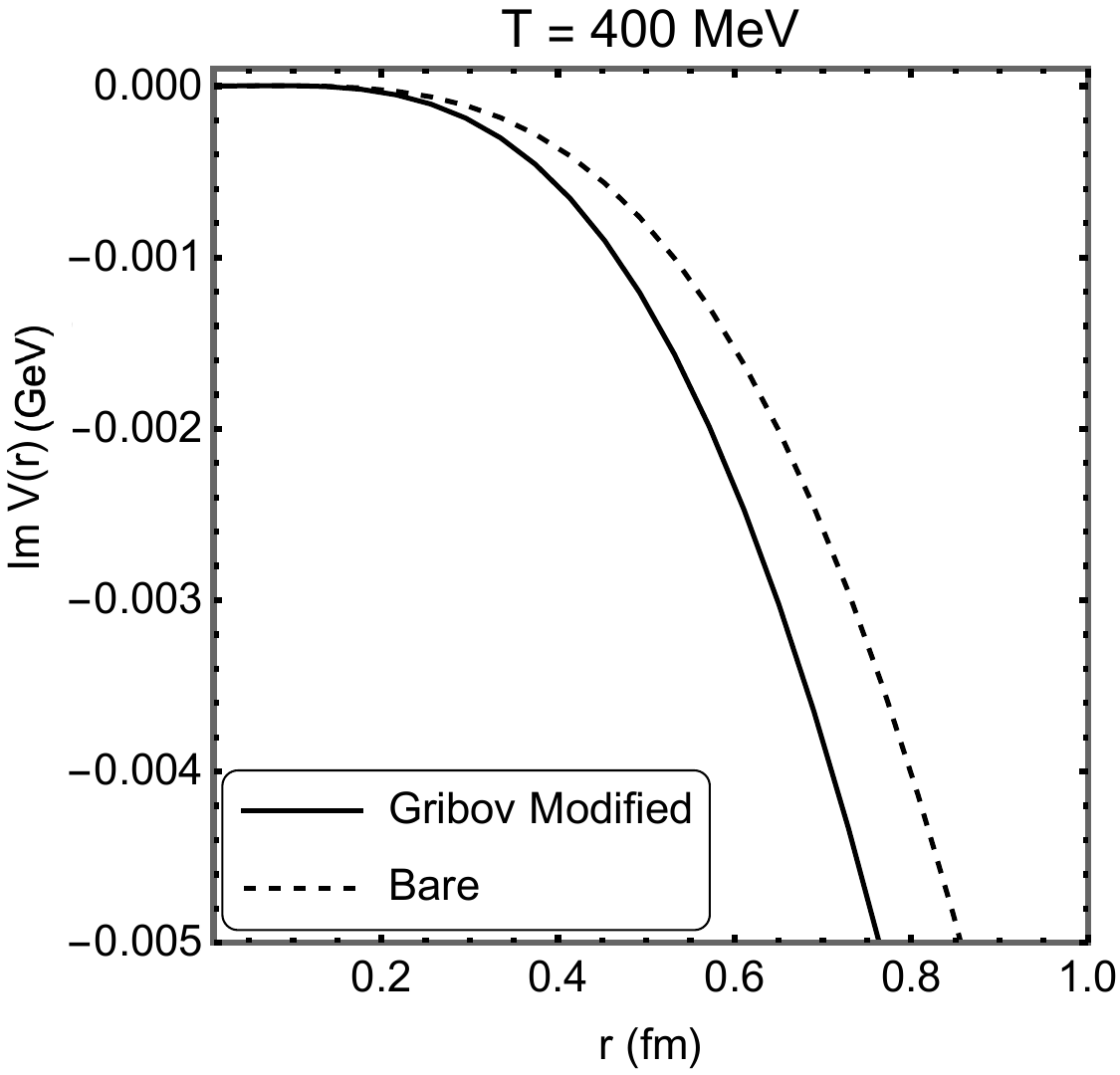}
    \caption{Comparison between the bare and the GZ modified imaginary part of the HQ potential while varying with distance.}
    \label{fig:ImV_gribov}
\end{figure*}

As an application of the Debye mass evaluated in section~\ref{sec2}, in this section we will estimate the corresponding modifications on the real and imaginary parts of the complex HQ potential. In general, the real part of the complex HQ potential corresponds to the screened Coulomb interaction and describes how the attractive force between the quark and antiquark is weakened by the medium. Hence it is directly relevant to the screening length or the Debye screening mass. On the other hand, the imaginary part of the HQ potential is closely related to Landau damping, where gluons mediating the quark-antiquark interaction are absorbed or scattered by thermal particles in the plasma. This leads to a dissipation of the interaction energy, causing quarkonium states to have a finite lifetime and ultimately dissociate at high temperatures. This phenomenon is a key aspect of understanding quarkonium suppression in heavy ion collision experiments, providing insight into the properties of the quark gluon plasma. Considering the full Cornell potential $V(r) = -\frac{\alpha}{r}+\sigma r$, containing both the Coulomb and the string part, the real and imaginary parts of the HQ potential is dependent on the temporal component of the retarded propagator and hence can be obtained in terms of the Debye mass. The simplified versions of the same are given in a modified coordinate space $\hat{r}=rm_D$ by taking the short-distance limit $ \hat{r} \ll 1$ as~\cite{Burnier:2009yu,Agotiya:2016bqr,Nilima:2022tmz} :   
\begin{eqnarray}
	\label{pis}
{\rm Re}\, V(\hat{r},T)&=&\left(\frac{2\sigma}{m_D}-\alpha\, m_D\right)\frac{e^{-\hat r}}{\hat r}\nonumber\\
	&-&\frac{2\sigma}{m_D \,\hat r}+\frac{2\sigma}{m_D}-\alpha \,m_D~,\label{rHQp}\\
	{\rm Im}\,V (\hat{r},T)&=&T\left(\frac{\alpha \,{\hat r^2}}{3}
	-\frac{\sigma \,{\hat r}^4}{30\, m_D^2}\right)\log\left(\frac{1}{\hat r}\right).
	\label{iHQp}
\end{eqnarray}
As one can see from Eqs.~\eqref{rHQp} and \eqref{iHQp}, in this simplified approximation the non-perturbative contribution through the GZ action in the heavy quark potential is being incorporated through the Debye mass $m_D$. As both the real and imaginary parts of the complex HQ potential are acutely dependent on the medium modified Debye screening mass $m_D$, any modification within $m_D$ will be reflected in the HQ potential. The long distance suppression of gluonic modes by the infrared regulation provided within the GZ framework will alter the behavior of the real and imaginary parts of the HQ potential as we will see in Figs.~\ref{fig:ReV_gribov} and \ref{fig:ImV_gribov}. At this point we again emphasise that all these physical explanations regarding the real and imaginary parts of HQ potential come into play for a QCD medium. Here, for an academic exercise we are instead using the EM Debye mass.

In Figs.~\ref{fig:ReV_gribov} and \ref{fig:ImV_gribov} we show the variations of respectively the real and imaginary parts of the HQ potential with respect to distance. For both the cases we have chosen two different temperatures, one relatively lower at $T=200$ MeV and the other in the higher temperature regime at $T=400$ MeV. It can be observed that for higher temperature the modification in the HQ potential due to the GZ modified quark propagation is more prominent. This is aligned with the fact that at high temperatures the interplay between the excluded Landau cut contributions and Gribov mode contribution becomes significant within the GZ framework, mimicing the results from section~\ref{sec2}. Furthermore, one can explicitly notice from both Figs.~\ref{fig:ReV_gribov} and \ref{fig:ImV_gribov}, that for larger values of temperature, the Gribov modified result diverges from the bare result rapidly, i.e. at a lower value of the distance $r$. This signifies that the long distance suppression incorporated in the HQ potential through the GZ modified Debye screening mass, becomes more dominant with increasing temperature. The change in the real part of the potential is more visible in this context reflecting its direct connection with the screening length. It is worth noting that in this work, we have just focused on the modifications of the HQ potential, and have not explored other quantities related to HQ potential, e.g. binding energy or dissociation temperature. However, through these preliminary observations we can rightfully conclude that once a more complete study of QCD Debye mass incorporating the GZ modified quarks sees the light of the day, it would be worth exploring the physical quantities such as HQ dissociation, which depends on the HQ potential.

\section{Summary and Outlook}
\label{sec4}

The present study explores the electromagnetic Debye mass as an important signature of quark matter, by incorporating the GZ modified quark propagator in the medium. This spontaneously includes a novel Gribov mode into the collective excitations of the quark matter which gets reflected in the corresponding observables computed, i.e. EM Debye mass and HQ potential. Absence of the Landau cut in the GZ modified quark spectral function means there are no explicit high temperature contributions from the quasigluon sector in the present scenario. However the spacelike Gribov mode compensates for this and encompasses medium effects up to considerably high momenta for higher temperatures. As an evidence of this fact, in the case of the EM Debye mass we find that with increasing temperature, the GZ modified Debye mass starts to dominate its bare counterpart due to the role played by the increasing contributions from the Gribov and plasmino mode. This enhanced EM Debye mass is expected to influence further prominent effects in corresponding dynamic signatures of quark matter, e.g. quarkonia suppression. In view of this we have also provided an estimation about the modifications on the HQ potential, albeit in a QED medium as an application of the GZ modified EM Debye mass. 

To the best of our knowledge, this is the first effort towards evaluating the Debye mass within the GZ action. To get a first qualitative estimate about the modification of the Debye mass, we have chosen to evaluate the EM Debye mass. Naturally the next step, which would serve better practical purposes, will be to explore the QCD Debye mass. For that case we need to compute more number of diagrams due to the self coupling of gluon which requires some nontrivial effort. It will be quite interesting to see how the QCD Debye mass with GZ modified quarks compares with and excels the previously obtained non-perturbative results of the same that include the $\mathcal{O}(g^2T)$ corrections~\cite{Arnold:1995bh,Kajantie:1997pd}. This would also help us to accurately estimate the HQ potential in a QCD medium and corresponding other relevant physical observable. Also for our present preliminary study, we have only chosen the asymptotic value of the Gribov parameter $\gamma_G$, solved from the gap equation. However, some recent studies~\cite{Fukushima:2013xsa, Jaiswal:2020qmj,Madni:2022bea} have also used modified versions of the Gribov parameter, which can be more appropriate to capture the physics for relatively lower values of temperature. These modifications would also provide some interesting takes in calculating the Debye Mass more precisely.

Finally we have to address that while the GZ framework has made significant strides in understanding confinement and the infrared behavior of QCD, it faces several theoretical and practical challenges. One of the main issue is the non-locality of the action which restricts gauge configurations to the Gribov region and makes the role of Gribov copies beyond the first Gribov horizon unclear. This limits the ability to rigorously test the predictions of the GZ framework, as non-local actions are difficult to treat in lattice QCD calculations, which are commonly employed to study non-perturbative aspects of gauge theories. Furthermore while the GZ framework suggests gluon confinement through the suppression of the gluon propagator in the infrared regime, there are other competing mechanisms for confinement, such as center vortex models, monopole condensation, and dual superconductor models, which are not well understood through GZ action. Hence it offers a partial picture of confinement instead of a fully comprehensive explanation. Also GZ framework does not directly explain how confined gluons and quarks combine to form observable hadrons, leaving a gap between theoretical predictions and experimental data in QCD. Further refinement (like the refined GZ approach) and exploration of alternative models may be necessary to resolve these open questions in modern gauge theory research.

\section*{Acknowledgements}

The author would like to thank Sanjukta Paul and Arghya Mukherjee for helpful discussions. This work was supported by the postdoctoral research fellowship of Alexander von Humboldt Foundation.

\bibliographystyle{unsrt}

\bibliography{mD_gribov}

\end{document}